\documentclass[a4paper,11pt]{article}
\usepackage{aaskaiid}
\usepackage{orcidlink} 
\usepackage{xspace}
\usepackage{bm}

\newcommand{\farcs}[1]{$.\!\!^{\prime\prime}${#1}}
\newcommand{\farcm}[1]{$.\!\!^{\prime}${#1}}
\newcommand{\HI}{\texorpdfstring{H\,{\sc i}}{H I}\xspace}
\newcommand{\HII}{\texorpdfstring{H\,{\sc ii}}{H II}\xspace}

\title{Measuring Magnetic Fields Near and Far with the SKA via the Zeeman Effect}
\ShortTitle{Zeeman Measurements Near and Far}
\ShortName{Robishaw et al.} 

\author[1,2]{Timothy Robishaw\,\orcidlink{0000-0002-4217-5138}}
\affiliation[1]{Dominion Radio Astrophysical Observatory, Herzberg Astronomy \& Astrophysics Research Centre, National Research Council Canada, 717 White Lake Rd., Kaleden, BC, V0H 1K0 Canada}
\affiliation[2]{Department of Computer Science, Mathematics, Physics \& Statistics, University of British Columbia, Okanagan Campus, Kelowna, BC V1V 1V7, Canada}
\emailAdd{tim.robishaw@nrc-cnrc.gc.ca}

\author[3,4]{Marta Nowotka\,\orcidlink{0009-0002-0282-4188}}
\affiliation[3]{Department of Physics, Stanford University, Stanford, CA 94305, USA}
\affiliation[4]{Kavli Institute for Particle Astrophysics \& Cosmology, P.O. Box 2450, Stanford University, Stanford, CA 94305, USA}

\author[5]{Tao-Chung Ching\,\orcidlink{0000-0001-8516-2532}}
\affiliation[5]{Department of Astronomy, Tsinghua University, Beijing 100084, People’s Republic of China}

\author[6]{James A. Green\,\orcidlink{0000-0002-2670-188X}}
\affiliation[6]{SKA Observatory, SKA-Low Science Operations Centre, ARRC Building, 26 Dick Perry Avenue, Technology Park, Kensington, 6151, WA, Australia}

\author[7]{Anita M.~S. Richards\,\orcidlink{0000-0002-3880-2450}}
\affiliation[7]{Jodrell Bank Centre for Astrophysics, School of Physics and Astronomy, Univ. of Manchester, M13 9PL, UK}

\author[7]{Sandra Etoka\,\orcidlink{0000-0003-3483-6212}}

\author[8]{Malcolm Gray\,\orcidlink{0000-0002-2094-846X}}
\affiliation[8]{National Astronomical Research Institute of Thailand, 260 Moo 4, T. Donkaew, A. Maerim, Chiangmai 50180, Thailand}

\author[3,4]{Susan E. Clark\,\orcidlink{0000-0002-7633-3376}}

\author[9]{Tyler L. Bourke\,\orcidlink{0000-0001-7491-0048}}
\affiliation[9]{SKA Observatory, Jodrell Bank, Lower Withington, SK11 9FT, UK}

\author[10]{Vincent Fish\,\orcidlink{0000-0002-7128-9345} }
\affiliation[10]{MIT Haystack Observatory, 99 Millstone Road, Westford, MA 01886, USA}

\abstract{Zeeman splitting in spectral lines---both in emission and absorption---provides direct estimates of magnetic field strength and direction in magnetized gas in our own Milky Way and in external galaxies. We discuss the potential for using the Square Kilometre Array (SKA) to measure the Zeeman effect in targets spanning an enormous range of distance: from cometary comas in the solar system, through Galactic molecular clouds, \HI filaments in the cold neutral medium, high-velocity clouds, the Fermi Bubbles, and photodissociation regions (PDRs) traced by radio recombination lines, to OH masers and megamasers in nearby and distant starburst galaxies, and to cold neutral gas in damped Ly$\alpha$ absorbing systems at cosmological redshifts. We update the sensitivity calculations of \citet{robishaw+15} and indicate, for each science goal, whether it will be achievable with Array Assembly~4 (AA4) of SKA-Mid, with the staged delivery of AA$^{*}$, or only with the full SKA buildout. Zeeman measurements will probe the magnetic field \textit{in situ} in the warm and cold neutral interstellar medium, complementing SKA Faraday rotation programs; radio recombination lines, stackable across hundreds of transitions, extend this reach to \HII regions and PDRs. In external galaxies, SKA-Mid will enable Zeeman studies of OH kilomasers in nearby starburst systems, substantially expand the census of megamaser Zeeman detections beyond the Arecibo sky, and probe magnetic fields in damped Ly$\alpha$ systems to field limits well below those currently achievable, opening a new window on the role of magnetic fields in galaxy formation and cosmic evolution.}

\begin{document}

\maketitle


\section{Introduction}

The editors and the SKA Cosmic Magnetism Working Group have asked us to present again the \citet{robishaw+15} chapter from \textit{Advancing Astrophysics with the Square Kilometre Array} (AASKA14), this time updated with new sensitivity calculations and recent developments from the past decade.

\citet[][this volume]{bourke+26} provide a thorough introduction to the Zeeman effect and its measurement.  A succinct overview follows here.

The Zeeman effect leaves its fingerprint on the circular polarization of a spectral line.  For certain atomic and molecular transitions, a spectral line emitted or absorbed in magnetized gas will exhibit circular polarization in which the two senses ($\sigma$ components) are split apart in frequency.  The amplitude of the observed frequency splitting between the $\sigma$ components is proportional to the total magnetic field scaled by a constant that depends on the atomic or molecular transition:\footnote{We incorrectly stated in \citet{robishaw+15}, eq.~(1), that the frequency splitting was proportional to the line-of-sight field component $B_\mathrm{los}$. The splitting is always proportional to the total field strength $B$; it is only the amplitude of the Stokes~$V$ $\mathsf{S}$-curve in the narrow-splitting regime that yields $B_\mathrm{los}$ via the $\cos\theta$ projection. In the resolved case, the total magnitude $B$ is recovered free of this projection, but the inclination of the field to the line-of-sight remains undetermined \citep[see][\S\S~2.3--2.4]{robishaw08}.}
\begin{equation}\label{eq:zeeman_splitting}
\Delta\nu[\mathrm{Hz}] = b B[\mu\mathrm{G}] (1+z)^{-1}\,,
\end{equation}
where $b$ is the transition-dependent Zeeman coefficient in Hz~$\mu$G$^{-1}$ (sometimes denoted $Z$ in the literature), $B$ is the magnetic field measured in microgauss, and $z$ is the redshift of the source; tables of all known Zeeman coefficients can be found in \citet{heiles+93}, \citet{robishaw08}, and \citet{crutcherk19}.

In the narrow-splitting regime (Zeeman splitting $\Delta\nu \ll$ Doppler line width), the two $\sigma$ components are unresolved and the net circular polarization spectrum, Stokes~$V \equiv \mathrm{RCP} - \mathrm{LCP}$, takes the form of an $\mathsf{S}$-curve whose amplitude is proportional to the derivative of the total intensity Stokes~$I$ profile. Here RCP and LCP follow the IEEE convention \citep{ieee18}, in which right-hand circular polarization denotes counterclockwise rotation of the electric-field vector as seen by the observer (opposite to the convention used in physics and optical astronomy); Stokes~$V$ as defined above follows the IAU convention \citep{iau74}. Because the $\sigma$ components carry opposite circular polarizations with intensities weighted by $\cos\theta$, where $\theta$ is the angle between the magnetic field and the line of sight, only the line-of-sight component $B_\mathrm{los} = B\cos\theta$ is recoverable from a least-squares fit of the derivative of Stokes~$I$ to the Stokes~$V$ spectrum.

The amplitude of the effect measured in the Stokes~$V$ spectrum toward sources in the Galactic interstellar medium (ISM) is generally very weak compared to the total intensity of the line measured in the Stokes~$I$ spectrum because the splitting is usually very small compared to the width of the line.  However, some Galactic maser emission lines can be so narrow that the two circularly polarized $\sigma$ components are completely resolved; in this case the observed frequency separation yields the total field strength $B$ directly---free of the $\cos\theta$ ambiguity inherent in the narrow-splitting regime---via eq.~(\ref{eq:zeeman_splitting}), and the sense of the circular polarization of each component determines the field direction along the line of sight \citep{crutcher+93}.

In this chapter, we discuss the potential for using SKA-Mid to measure the Zeeman effect in various targets of interest in roughly increasing order of distance.


\subsection{Assumptions in This Chapter}

In this chapter, we update all observing time estimates from \citet{robishaw+15} and indicate, for each Zeeman science goal, whether it will be achievable with Array Assembly~4 (AA4) of SKA-Mid, will be achievable with the reduced capabilities provided by the staged delivery of AA$^{*}$, or conversely will require a full SKA buildout.

Table~\ref{tab:ska_sensitivity} shows our AA$^{*}$ and AA4 sensitivity estimates for common Zeeman transitions in Bands~2 (\mbox{0.95--1.76~GHz}) and~5a (\mbox{4.6--8.5~GHz}) of SKA-Mid.  We provide estimates assuming Briggs weighting for two values of the robustness parameter, $R=\pm1$. We assume a source at declination $-20^{\circ}$ and a telescope elevation of $45^{\circ}$.  We round the tabulated sensitivity estimates because they will vary with elevation as a source is tracked.  The SKA2 sensitivity estimates in \citet{robishaw+15} reflected the specifications available at that time; the full SKA vision (formerly SKA2) is now expected to deliver tenfold greater sensitivity and twentyfold finer angular resolution than AA4 (formerly SKA1).




\begin{table}
\caption{SKA-Mid sensitivities from \protect\url{https://sensitivity-calculator.skao.int/mid} for common Zeeman transitions in Bands~2 and~5a. $R$ is the robustness parameter for Briggs weighting. Continuum sensitivities are computed over the full listed bandwidth without excluding spectral lines. Velocities are spectral resolution. Sensitivity values are rms noise measured in $\mathrm{\mu Jy~bm^{-1}}$. Corresponding beams are given below each sensitivity estimate.  Continuum estimates are not listed for the 18~cm OH and 6.6~GHz CH$_3$OH transitions as they are nearly identical to the 21~cm \HI and 6~GHz OH values, respectively. Source declination is set to $-20^{\circ}$, elevation is $45^{\circ}$. No tapering is applied. Estimates are rounded as actual values will vary with elevation.}
\label{tab:ska_sensitivity}
\begin{tabular}{l|cc|cc}
\hline
\HI 1.420 GHz (or high-$z$ OH) &\multicolumn{2}{c|}{AA$^*$ {\bf 1.42 GHz}} &
\multicolumn{2}{c}{AA4 {\bf 1.42 GHz}}\\
\hline
Integration Time: 1 h &
$R=+1$ & $R=-1$ &
$R=+1$ & $R=-1$ \\  
\hline
Continuum (0.5 GHz) rms $\left[\mathrm{\mu Jy~bm^{-1}}\right]$ &
1.5 &
4.3 &
1.0 &
2.2 \\
Continuum Beam  &
2\farcs8$\times$2\farcs5 & 
1\farcs0$\times$0\farcs8 &
1\farcs0$\times$0\farcs9 &
0\farcs4$\times$0\farcs4\\ 
0.35 km s$^{-1}$ rms $\left[\mathrm{\mu Jy~bm^{-1}}\right]$ &
790 &
2500 &
560 &
1430 \\
Spectral Line Beam &
3\farcs0$\times$2\farcs5 &
1\farcs3$\times$1\farcs0 &
1\farcs0$\times$1\farcs1 &
0\farcs4$\times$0\farcs4\\
\hline
\hline
OH 1.612/1.665/1.667/1.720 GHz &\multicolumn{2}{c|}{AA$^*$ {\bf 1.66 GHz}} &\multicolumn{2}{c}{AA4 {\bf 1.66 GHz}}\\
\hline
Integration Time: 1 h &
$R=+1$ &$R=-1$ & 
$R=+1$ &$R=-1$ \\  
\hline 
0.15 km s$^{-1}$ rms $\left[\mathrm{\mu Jy~bm^{-1}}\right]$ & 
1100 & 
3400 & 
770 & 
1950\\
Spectral Line Beam  & 
2\farcs6$\times$2\farcs2 &
1\farcs1$\times$0\farcs9 & 
0\farcs9$\times$0\farcs9 & 
0\farcs4$\times$0\farcs4 \\
\hline
\hline
OH 6.016/6.031/6.035/6.049 GHz &\multicolumn{2}{c|}{AA$^*$ {\bf 6.04 GHz}} &\multicolumn{2}{c}{AA4 {\bf 6.04 GHz}}\\
\hline
Integration Time: 1 h &
$R=+1$ & $R=-1$ &
$R=+1$ & $R=-1$ \\
\hline
Continuum (2 GHz) rms $\left[\mathrm{\mu Jy~bm^{-1}}\right]$ &
1.3 &
3.4 & 
0.8 & 
1.6\\
Continuum Beam &
0\farcs51$\times$0\farcs45 &
0\farcs20$\times$0\farcs15 &
0\farcs19$\times$0\farcs19 &
0\farcs08$\times$0\farcs08\\ 
0.083 km s$^{-1}$ rms $\left[\mathrm{\mu Jy~bm^{-1}}\right]$ &
1400 &
4000 &
840 &
2000 \\
Spectral Line Beam &
0\farcs56$\times$0\farcs46 &
0\farcs27$\times$0\farcs22 &
0\farcs20$\times$0\farcs20 &
0\farcs09$\times$0\farcs09\\ 
\hline
\hline
CH$_{3}$OH 6.668 GHz &\multicolumn{2}{c|}{AA$^*$ {\bf 6.67 GHz}} &\multicolumn{2}{c}{AA4 {\bf 6.67 GHz}}\\
\hline
Integration Time: 10 h &
$R=+1$ & $R=-1$ &
$R=+1$ & $R=-1$ \\  
\hline
0.009 km s$^{-1}$ rms $\left[\mathrm{\mu Jy~bm^{-1}}\right]$ &
1300 &
3700 &
780 &
1860\\
Spectral Line Beam &
0\farcs50$\times$0\farcs42 &
0\farcs25$\times$0\farcs20 &
0\farcs18$\times$0\farcs18 &
0\farcs08$\times$0\farcs08\\ 
\hline
\end{tabular}
\end{table}


\section{Cometary Magnetic Fields}

OH masers in comets offer a means of measuring coma magnetic fields through Zeeman splitting. At heliocentric distances of 1 to a few~au, comet outgassing produces beam-averaged OH column densities of order 10$^{19}$~m$^{-2}$ \citep{smith+21} in the 18~cm lines, with the masing coma subtending a radius of 1$^{\prime}$--$\,\sim$10$^{\prime}$ around the nucleus \citep{colom+11} and an expansion velocity of order 1~km~s$^{-1}$. The maser is stimulated by a solar UV line and may appear in absorption (usually against the Galactic background) or emission depending on the relative velocity vectors of the comet with respect to the Sun and observer \citep[the Swings effect;][]{despois+81}. The main lines (1665, 1667~MHz) are strongest, although the satellite lines (1612, 1720~MHz) are also detected. Zeeman splitting has been measured around a few comets, giving magnetic field strengths of a few tens to a few hundred microgauss \citep{gerard85, gerard+98}; depending on the comet location, this measurement reveals details of the solar/interplanetary magnetic field and its relationship with the cometary field. Connecting these measurements to the ambient interplanetary field requires modeling the dynamic interaction of ions and neutral molecules in the coma, informed by ionized-tail observations \citep{gan-baruch+98, wellbrockj25} and Atacama Large Millimeter/Submillimeter Array (ALMA) mapping of the neutral molecule distribution \citep{cordiner+14, roth+21, cordiner+23}.

Most cometary OH observations have been made with single dishes at typical resolutions of 3$^{\prime}$--$\,\sim$10$^{\prime}$, sometimes with limited mosaics. Zeeman splitting of OH main lines has been measured using Nan\c{c}ay observations with a typical resolution of 3\farcm5$\times$19$^{\prime}$ in 0.28~km~s$^{-1}$ channels at a sensitivity of $\sim$20~mJy ($\sim$37~mK if the comet fills the beam), requiring spectral fitting to recover the RR--LL spectral offset of order 0.05~km~s$^{-1}$.  More accurate Zeeman-splitting measurements would be obtained using AA4 in 10~h, with visibilities tapered to a resolution of 3\farcm5 in 0.04~km~s$^{-1}$ channels, yielding an rms noise of $\sim$2.5~mJy~beam$^{-1}$ (30~mK).  Using AA$^{*}$, a similar brightness-temperature sensitivity is reached at a resolution of $\sim$7$^{\prime}$ and channel width of 0.15~km~s$^{-1}$, still providing more detail than previously achieved.  Furthermore, if data suitable for reimaging are supplied, different channel averaging or image weighting can improve sensitivity or resolution as needed---for example, to detect the weaker satellite lines (not yet detected in polarization), to locate the brightest lines with sub-arcminute accuracy, or to test models for the radii of OH formation and destruction.


\section{Molecular Clouds and Star-Forming Regions in the Milky Way}

Material from this section of \citet{robishaw+15} covering the thermal-line Zeeman effect in star-forming regions is treated in depth in the chapter ``Measuring Magnetic Field Strengths in Galactic Star-Forming Regions via the Zeeman Effect'' by \citet{bourke+26} in this volume; \citet{bourke+26} defer maser emission from evolved stars and regions of high-mass star formation (HMSF) to the current chapter and \citet{rygl+26}.

SKA-Mid will be invaluable in the study of Galactic masers \citep{rygl+26}, including very high-resolution follow-up of catalogs such as the Methanol (and OH) Multibeam Survey \citep{green+09}. Multiple maser transitions provide a chronometer with thousand-year resolution for HMSF concentrated in the Galactic plane \citep{ellingsen07,breen+10}. Southern Hemisphere centimeter-wave studies with resolutions of a few hundred milliarcseconds or better are needed to complement ALMA and to map magnetic field structures. At the other end of stellar evolution, we do not fully understand the precise mechanisms by which cool red giants, AGB stars, and supergiants lose mass, or how nearly spherical stars give rise to asymmetric planetary nebulae (PNe) and supernova remnants; masers serve as uniquely powerful probes of both.  SKA-Mid and ALMA will not only map the kinematics and magnetic fields at high resolution with masers, but also, for the first time, relate these to thermal lines and dust emission with high precision.

Although O-rich evolved stars begin their lives as spherically symmetric objects, a large fraction of PNe exhibit asymmetric morphologies. The trigger for the asymmetry of PNe is most likely binarity, but the organized magnetic fields detected in the regions where OH masers operate in the circumstellar envelopes (CSEs) indicate a possible role of magnetism in this morphological development. These evolved stars generally exhibit maser emission from three out of the four possible ground-state 18~cm OH transitions, namely the two main-line transitions at 1665 and 1667~MHz and the 1612~MHz satellite transition; all three generally show some degree of linear and circular polarization. The maser emissions emanate from different locations and depths in the CSE with the main-line transitions generally arising deeper within the CSE than the 1612~MHz transition, which probes the outermost part of the CSE. The magnetic field strength and structure at the location of these maser emissions can therefore be retrieved for a wide range of evolved stars (on the AGB track but also for the red supergiants) and evolutionary stages \citep{etokad04,etokad10,etoka+17}, through to the proto-PN stage \citep{etoka+09}. Thousands of such objects are expected to be detectable in these transitions throughout the Galaxy down to a flux limit of 4~mJy \citep{etoka+15}; for 0.6~km~s$^{-1}$ channels, SKA-Mid AA4 would reach this sensitivity in 4~min toward the Galactic center, with Briggs weighting and $R=+1$ yielding a survey-mode beam size of 1$^{\prime}$$\times$1$^{\prime}$ (10~min with a beam of 3$^{\prime}$$\times$2$^{\prime}$ for AA$^{*}$).

OH maser emission at 1720~MHz is not observed toward standard AGB stars because the conditions for its inversion are not favorable. On the other hand, it has been detected toward young PNe \citep{gomez+16,qiao+16}. Similarly, excited-OH maser emission at 6~GHz has been detected toward PNe \citep{desmurs+10,houg20}, and recently also at 4.6~GHz toward a so-called water-fountain source \citep[thought to be a transitional phase between the AGB and proto-PN stage;][]{ouyang+24}. Zeeman signatures have been found in these transitions, making them an additional potential probe of the magnetic field in these types of very evolved and short-lived objects.


\section{The Magnetic Field Structure of the Milky Way}

The Zeeman effect will be used to study the large-scale magnetic field structure throughout the Galaxy.  Unlike Faraday rotation measurements, the polarization signature of the Zeeman effect will not be perturbed by the intervening medium, probing instead the \textit{in situ} field strengths at the site of spectral-line emission or absorption. SKA-Mid Zeeman measurements will therefore complement the extensive Faraday rotation mapping programs described by \citet{heald+20} and in this volume by \citet{tahani+26}, \citet{sun+26}, and \citet{ma+26}.

\subsection{\HI Absorption Lines against Background Continuum Sources}

The magnetic field in the cold neutral medium (CNM) can be probed via the Zeeman effect in 21~cm absorption of continuum emission from compact background sources. This method was used by \citet{heilest04,heilest05} in their Millennium Arecibo 21~cm Absorption-Line Survey at the Arecibo 305~m telescope---supplemented by archival data from the Hat Creek Radio Observatory 85~ft telescope---to make an unprecedented survey of magnetic fields in the CNM toward 41 radio-loud sources. Gaussian decomposition of the 41 absorption spectra produced 136 distinct velocity components; for each, the velocity derivative of the fitted Stokes~$I$ component was scaled to the corresponding Stokes~$V$ spectrum in a least-squares fit for $B_{\mathrm{los}}$. Of the 136 components, 69 met the quality threshold $\sigma_{B_{\mathrm{los}}} < 10~\mathrm{\mu G}$, and 22 of those yielded detections at a signal-to-noise ratio $\mathrm{SNR}>2$. As the largest \HI absorption Zeeman survey to date, this data set has provided crucial constraints on the energy balance in the CNM \citep{heilest05} and the scaling of magnetic field with density, a key diagnostic of the role of magnetic fields in molecular cloud formation \citep{crutcher+10, tritsis+15, jiang+20, hu+23, setam25, whitworth+25}. The Millennium Survey sample has also enabled new insights into the magnetic field geometry: recently, \citet{nowotka+25} found a correlation between Zeeman measurements and the morphology of \HI emission filaments tracing the plane-of-sky magnetic field, suggesting that these tracers can be combined to constrain the total magnetic field strength. Because existing studies have relied almost entirely on the Arecibo data, deeper and more extensive Zeeman measurements in \HI absorption are crucial for further progress.

The limiting sensitivity of the Arecibo survey was 3~mJy.  For 0.35~km~s$^{-1}$ channels, SKA-Mid AA4 would achieve this sensitivity in 13~min (40~min for AA$^{*}$). Therefore, the first census of the CNM magnetic fields throughout the southern sky should be easily achievable. In the northern sky, the Five-hundred-meter Aperture Spherical radio Telescope (FAST) has already demonstrated its capability for this task by obtaining results consistent with Arecibo toward three radio sources \citep{nowotka+25}. Single-dish \HI absorption spectra are, however, produced by subtracting an off-position spectrum and are therefore prone to inaccuracies; the possibility also arises that polarized sidelobes can contaminate Stokes~$V$ spectra for weak sources when using this method. Both problems are mitigated by interferometric measurements. With its high sensitivity and access to a sky footprint previously unexplored in 21~cm polarization, SKA-Mid will enable a substantial expansion of magnetic-field measurements in cold neutral hydrogen.

\subsection{\HI Self-Absorption}

An intriguing group of targets includes positions that show 21~cm self-absorption \citep[the ``self'' prefix distinguishing this case from that of absorption against a background continuum source;][]{lig03}; for example, dozens of positions throughout the Perseus and Taurus molecular clouds show \HI self-absorption (HISA) at velocities coinciding with OH and CO emission. Zeeman observations of HISA have been obtained with detections of field strengths between 2.1 and 17.2~$\mu$G (with uncertainties ranging from 0.5 to 3.4~$\mu$G) in the Ophiuchus cloud \citep{goodmanh94} and a field strength of $3.8 \pm 0.3$~$\mu$G in the Taurus cloud \citep{ching+22}. These detections demonstrate the potential of HISA as a common probe of Galactic magnetic field structure.  \citet{mcclure-griffiths+06} detected HISA throughout the Riegel-Crutcher (R-C) cloud, an 8$^\circ$$\times$8$^\circ$ region at the Galactic center showing a striking network of long, cold filaments.  When the method of \citet{chandrasekharf53} is used to investigate stellar polarization measurements along these filaments, magnetic fields in excess of 30~$\mu$G are inferred.  The HISA profile in the R-C cloud has a depth of 100~K and a width of 3.5~km~s$^{-1}$.  A 30~$\mu$G field produces a Stokes~$V$ profile with peak-to-peak extent only 1.6\% of the line height, or 1.6~K.  Such a profile can be fitted if observed at a sensitivity of 0.3~K.  The width of the filaments is 2$'$--5$'$, so an integration time of 40~min per pointing would be required if observed by AA4 (2~h for AA$^{*}$) with a 60$''$ beam and 0.35~km~s$^{-1}$ channels; even AA$^{*}$ therefore suffices to demonstrate the utility of HISA Zeeman observations in the R-C cloud.  We can begin to map the field in filamentary structures with AA4. A complete map of the R-C cloud would require hundreds of pointings and is best deferred to the full SKA vision, whose sensitivity and field of view will exceed those of AA4 by factors of ten and twenty, respectively.

\subsection{Diffuse 21~cm Emission}

Diffuse 21~cm emission is seen in every direction in the sky.  This fact, combined with the strong splitting coefficient for the 21~cm line, makes \HI emission an enticing target for Zeeman studies.  This line of work was pursued successfully using single dishes for decades before being abandoned in the mid-1990s \citep{heilest82,heiles88,heiles89,heiles96,heiles97,trolandh82a,trolandh82b,verschuur89,goodmanh94,myersk95}.  Reliable measurements require a careful accounting of instrumental circular polarization contributions from polarized sidelobes of the telescope \citep{robishawh21}.

\subsubsection{Diffuse \HI Science Targets}

Zeeman observations of 21~cm emission in high-velocity clouds (HVCs) will reveal the large-scale magnetic field structures beyond the scale height of the Milky Way disk. Faraday rotation and dust polarization can probe the field within the disk but are less effective in the halo, leaving the disk--halo interface poorly constrained. HVCs travel through this interface, interacting with ambient gas and magnetic structures, and Zeeman observations will trace how the field threads through and confines these clouds and how the cloud motions in turn distort the field. The first HVC target will be the Smith Cloud, whose trajectory and 3-D space motion are well constrained \citep{fox+16}. Using Faraday rotation measures toward extragalactic sources behind the Smith Cloud, \citet{hill+13} made the first detection of magnetic fields in an HVC, and \citet{betti+19} derived a lower limit of 5~$\mu$G toward the surrounding ionized medium. Zeeman measurements of the Smith Cloud can reveal how the magnetic field governs its survival and clarify the cloud's origin. More broadly, Zeeman mapping of HVCs will help reconstruct the geometry, strength, and continuity of the Milky Way’s magnetic field from disk to halo and open a new window on the role of magnetic fields in regulating gas accretion onto the Galactic disk.

\HI clouds surrounding the Fermi Bubbles are valuable targets for Zeeman observations of diffuse 21~cm emission. The bubbles exhibit sharp edges, high-temperature plasma, and signatures of ordered magnetic structures; magnetic fields play a major role in shaping them, confining cosmic rays, and stabilizing their surfaces against hydrodynamic instabilities \citep{sarkar24}. Synchrotron studies suggest magnetic fields of a few microgauss in the bubble lobes \citep{carretti+13}, while models of magnetic draping and bubble expansion predict that fields can be compressed and amplified up to tens of microgauss along the bubble walls \citep{yang+13}.  Observationally, \HI studies reveal a strong anti-correlation between Galactic-center \HI emission and the $\gamma$-ray bubble structures, including large \HI voids surrounded by high-velocity neutral gas tracing the bubble boundaries \citep{lockmanm16}. Zeeman measurements toward these boundary \HI clouds therefore provide a localized probe of the magnetic field strength and geometry at the bubble edges, directly testing whether the fields have been amplified or reoriented by the expanding outflow. Such measurements would constrain whether the Fermi Bubbles were formed by a nuclear starburst \citep{crockera11} or an episode of AGN activity \citep{guom12}.

\subsubsection{Instrumental Polarization: Challenges and Mitigation}

Because diffuse Galactic 21~cm emission appears along every direction in the sky, the large-scale, off-axis instrumental polarization response of a radio telescope can convert this wide-field total-intensity signal into an on-axis circularly polarized response \citep{robishawh21}.  To our knowledge, no one has successfully used an interferometer to measure 21~cm Zeeman splitting in diffuse emission. This scenario obtains at the Karl G.~Jansky Very Large Array (VLA) because the Cassegrain optics of each telescope, with the $L$-band receiver laterally offset from the symmetry axis, induce a severe beam squint in the circular polarization response \citep{jagannathan+17}.  This beam squint is not impossible to model, measure, and account for, but doing so is a major undertaking.  The SKA-Mid dishes, by contrast, employ an offset Gregorian design whose secondary reflector and feed alignment satisfy the ``Mizuguchi--Dragone condition'' \citep{mizuguchi+78,dragone78}, which should minimize cross polarization and, therefore, the associated Stokes~$V$ squint.  If the circularly polarized beam can be characterized for SKA-Mid, then we anticipate that the instrumental contribution to a Stokes~$V$ measurement can be modeled and accounted for---the reward will extend well beyond Zeeman science: \citet{agudo+15} claimed that relativistic jet studies require circular polarization precision of 0.01\%, a demanding standard that a fully characterized beam would satisfy.

It is therefore essential to model and measure the polarized beam pattern of any telescope performing Zeeman measurements using diffuse 21~cm emission.  A radio telescope modeling effort is underway at the Dominion Radio Astrophysical Observatory \citep{du+16,islam+24}.  Astronomers are working with mechanical and electrical engineers to study how telescope structures flex and their optics misalign as a function of telescope pointing positions; electromagnetic modeling software is then used to produce pointing-dependent polarized beam patterns.  These patterns can then be used to estimate how the telescope converts wide-field 21~cm emission into an on-axis instrumentally generated response that can mimic the Zeeman effect.  \citet{chen+25} recently demonstrated the utility of this modeling by comparing the expected polarized beam patterns for FAST's multi-horn receiver with the measured patterns, showing excellent agreement.  Full-Stokes beam modeling of this kind will be an essential part of any diffuse-emission Zeeman program, whether on a single-dish telescope or array, and we expect the results of this work to be directly applicable to SKA-Mid measurements of the Zeeman effect in 21~cm emission.  The many polarization projects planned for SKA-Mid will drive a thorough characterization of the polarization properties of the array, directly benefiting diffuse Zeeman science. We therefore expect SKA-Mid to be the first interferometer to seriously pursue this measurement.

\subsection{Radio Recombination Lines}

An exciting possibility for Zeeman studies in the ionized ISM of the Milky Way exists via the study of radio recombination lines \citep[RRLs;][]{oonk+15,salas+26}.  There are hundreds of hydrogen, helium, and carbon $n\alpha$ and $n\beta$ RRLs available across Bands~1 and~2.  Each of these RRL transitions is susceptible to the Zeeman effect with the same Zeeman coefficient as the 21~cm \HI line \citep[$b=2.8$~Hz~$\mu$G$^{-1}$;][]{grevep80}.  These RRLs can be stacked to gain sensitivity, and SKA-Mid could be used to study the magnetic fields threading Galactic \HII regions \citep{linville+23} and the diffuse ionized ISM in the Milky Way \citep{anderson+21,balser22}.  SKA-Mid would provide the spatial resolution to study photodissociation regions (PDRs) where carbon densities are enhanced and magnetic fields are expected to be amplified via flux-freezing; indeed, \citet{balser+16} have estimated magnetic field strengths of order 300~$\mu$G by studying the non-thermal line widths of carbon RRLs in PDRs.

\subsection{OH Masers Tracing the Galactic Magnetic Field}

\citet{davies74} found the remarkable result that the magnetic field measured in OH masers seemingly traces the large-scale field in which the masers are embedded. Several authors subsequently investigated this concept \citep[e.g.,][]{reids90,fish+03,hanz07}, finding fields coherent across kiloparsec scales. High-resolution VLBI observations demonstrate largely coherent field directions and magnitudes \citep[e.g.,][]{fish+03,vlemmingsl07}, a result replicated in lower-resolution single-dish studies \citep{fish+05,szymczakg09}. These studies were conducted largely with samples of masers collated from a range of heterogeneous observations; the largest set of systematic observations was that of \citet{fish+03}, but it was limited to only 40 star-forming regions, all visible from the Northern Hemisphere, with only a few masers sampled per spiral arm. Despite these limitations, the magnetic fields traced by the masers appear to be tied to the large-scale Galactic magnetic field, as traced by rotation measures \citep[e.g.,][]{brown+07,vaneck+11}.

The MAGMO survey \citep{green+12,green+15c,ogbodo+20} was launched at the Australia Telescope Compact Array (ATCA) to map the magnetic field in the Galactic plane by searching for Zeeman splitting in OH masers. It achieved a sensitivity of 50~mJy in 30~min of ATCA observations with an 8$''$ beam, and consisted of follow-up observations of positions where 6.7~GHz methanol masers were detected, finding field strengths up to $\sim$10~mG. As an exclusive tracer of HMSF \citep{minier+03,pestalozzi+05,xu+08,breen+13}, this maser species marks the key structural features of the Galaxy---the spiral arms, 3~kpc arms, and bar interaction. The combination of structure and magnetic field information can be a very powerful tool for understanding the dynamics and evolution of the Milky Way. Initial results of MAGMO indicate magnetic field coherence on larger scales than previously known, but with limited statistics; analysis of the full data set is required.

The logical extension of the MAGMO project would be a blind gridded survey for OH masers in the Galactic plane using SKA-Mid (see also \citealt{rygl+26} in this volume). A survey covering $-2^\circ$$<$$b$$<$$+2^\circ$ and $190^\circ$$<$$l$$<$$60^\circ$ (passing through the Galactic center) would subtend 920~deg$^2$, providing a sensitivity of 3.7~mJy at the 18~cm OH transitions with 0.04~km~s$^{-1}$ spectral resolution and a 10$''$ beam in 1.4~months of observing in the AA4 configuration with $R=+1$ (2.2~months using AA$^{*}$). The scale height for HMSF is well known and this latitude range will capture the vast majority of star-forming sources ($>$95\%).

A blind survey to better sensitivities than those achieved by the targeted MAGMO study will yield a higher detection rate toward regions of HMSF and will increase the number of Zeeman detections. This will substantially improve statistics, both within each region and within Galactic structures such as individual spiral arms. The increased number of measurements will enable a comparison of maser field directions with those probed via Faraday rotation of polarized continuum sources through nearby sightlines containing a magneto-ionic medium \citep{green+14}; the Faraday rotation grid itself will in turn be vastly improved in the SKA era through all-sky polarimetric surveys \citep{johnston-hollitt+15,heald+20,sun+26}.

An SKA-MAGMO study will explore whether the additional SKA-Mid sensitivity increases the detection rate toward HMSF regions, the number of sources detected per HMSF site, or both. With the extra sensitivity, we would expect an increase in the number of Zeeman pairs and triplets, which would provide more measurements to compare with Faraday rotation measures, decreasing the statistical error. The additional measurements will also allow exploration of the proportion of Zeeman pairs compared to Zeeman triplets for weaker sources, and thus the propensity of detectable $\pi$ components \citep[expanding on the analysis of][]{green+15a}. A statistically significant sample of $\pi$ components could provide an opportunity to supplement line-of-sight field studies with those in the plane.\footnote{The linearly polarized $\pi$ component, at the unsplit line frequency, yields both the strength and orientation of the plane-of-sky field $B_\perp$; see \citet[\S~2.4.1.3]{robishaw08}.} The SKA-MAGMO study would further enhance our understanding of the dynamics and evolution of our Galaxy through association with precise positions of HMSF regions \citep{green+15b}.


\section{Zeeman Splitting in External Galaxies}

Masers have the highest luminosity per unit frequency of any persistent radio source, so it is natural to use Zeeman splitting in these beacons to measure magnetic fields in distant galaxies.  Fortunately, as we saw above in discussing the MAGMO project, OH has a very large Zeeman coefficient and is therefore a sensitive tracer of magnetic fields.  We discuss the possibilities of using OH masers and megamasers as extragalactic magnetometers.

\subsection{OH Masers in Nearby Galaxies}

The full SKA could be used to survey the Local Group galaxies for OH maser emission and Zeeman splitting.  If detected, the field structure inferred would be of great interest when compared to MAGMO results in our own Galaxy because our Galactic study will be confined only to the plane.

\citet{brooksw97} detected OH masers in the LMC, needing a sensitivity of $\sim$40~mJy to obtain a 5$\sigma$ detection in Stokes~$I$.  For a channel resolution of 0.16~km~s$^{-1}$, AA4 would require 3~min to detect Zeeman splitting in such a source (10~min for AA$^{*}$).  A survey of the OH maser emission in the LMC and SMC is therefore of great interest: any field structure detected in the OH maser distribution could be compared directly with the fields already mapped in the LMC and SMC via Faraday rotation \citep{gaensler+05,mao+08}.

Although \citet{busch24} has recently used the Green Bank Telescope (GBT) to detect thermal 18~cm OH emission in the Andromeda Galaxy, OH masers have yet to be detected in M31.  \citet{willett11} conducted a search using the VLA, finding nothing above a threshold of 10~mJy at 5$\sigma$, and concluded that an order-of-magnitude increase in sensitivity would be required to probe the OH masers.  This would require 36~h of AA4 integration time with a velocity resolution of 0.62~km~s$^{-1}$, and an impractical 115~h with AA$^{*}$.  M31 is at declination $+40^{\circ}$, so SKA-Mid will only be able to track it over the elevation range $15^{\circ}$--$19^{\circ}$, making extended observations of M31 challenging but not infeasible.

M82 is the canonical starburst galaxy, located at a distance of 3.5~Mpc.  \citet{argo+10} used 61~h of VLA observing with a 1$''$ beam to show that OH masers (actually, \emph{kilomasers}) are seen throughout M82 with line widths of $\sim$10~km~s$^{-1}$ and fluxes as weak as 2~mJy.  M82 is at declination $+69^{\circ}$ and will never be observable by SKA-Mid, but it can serve as a proxy for the possibility of mapping OH masers in nearby starburst galaxies below declination $+40^{\circ}$, e.g., NGC~253, NGC~4945, and NGC~5253.  For a 5$\sigma$ detection of all the Stokes~$I$ features in M82, we would require 4.2~h of AA4 (13~h of AA$^{*}$) observation time with 1.2~km~s$^{-1}$ channels.  However, unlike in the Milky Way, the OH maser lines in M82 are broad enough that they are only partially split even for large fields of 3~mG.  The peak-to-peak amplitude of the Zeeman feature in Stokes~$V$ will be 50\% of the line height for a splitting induced by a 3~mG field. For the weakest detected OH maser, we can easily fit a Zeeman profile to a sensitivity limit of 0.3~mJy; detecting Zeeman splitting in all observed maser features using AA4 (and assuming a declination of $-25^{\circ}$, as appropriate for NGC~253) would therefore require 5~h of integration (16~h for AA$^{*}$).  This could allow for a mapping of the magnetic field as traced by OH masers in nearby starburst galaxies.

\subsection{Megamasers}

Megamasers have been detected in hundreds of galaxies at $z \sim 0.1$, and, with the aid of lensing, as far away as $z=2.6$ \citep{castangia+11}. Intrinsically compact, both spatially (detectable down to microarcsecond scales) and spectrally (sub-km~s$^{-1}$), they provide the best directly mappable tracers of high-resolution structure in the inner few hundred parsecs of active galaxies. SKA-Mid will be suited to studies of the larger-scale OH masers (rest frequency 1.67~GHz) associated with nuclear starbursts and Seyfert galaxies. The position of individual maser spots can be measured with an accuracy proportional to the ratio of beam size to SNR. For example, at $z \sim 0.05$, the Multi-Element Radio Linked Interferometer Network (MERLIN; resolution 120~mas) could resolve 10~pc details in Mrk~231 and Mrk~273 \citep{yates+00,richards+05} showing warped disks and orbiting mass densities of 300--900~M$_\odot$~pc$^{-3}$.

The most compelling use of OH megamasers (OHMs) has been the measurement of \textit{in situ} magnetic fields in 15 external galaxies by \citet{robishaw+08} and \citet{mcbrideh13}, who used the high spectral resolution of Arecibo to separate multiple Zeeman components and compared these with VLBI and MERLIN total-intensity lower-resolution spectra extracted from spatially discrete regions. The inferred magnetic field strengths of 0.5--80~mG provide an energy density comparable to the hydrostatic gas pressure in the masing regions, which are likely active star formation sites.  The results from OHMs also suggest that magnetic fields are dynamically important throughout the central starburst region, whereas radio synchrotron measurements that assume equipartition yield weaker inferred fields. The OHM-derived estimates are consistent with the linearity of the far-infrared--radio correlation \citep{mcbride+14}.  The only progress in Zeeman studies of OHMs since the 2015 version of this chapter \citep{robishaw+15} comes from \citet{mcbride+15}, who used the High-Sensitivity Array (HSA)\footnote{The HSA at the time comprised the Very Long Baseline Array (VLBA), the GBT, Arecibo, and the phased VLA.} to produce VLBI imaging of the Zeeman effect in Arp~220, providing spatially resolved confirmation of the single-dish Arecibo detection.  While two separate groups have used FAST to attempt confirmation of previous Zeeman detections in OHMs (neither has yet published results), only three of those galaxies are redshifted into the FAST observing band (whose upper limit is 1450~MHz).  The available sky coverage for FAST, moreover, is only moderately larger than Arecibo's was, and well short of that of SKA-Mid.

The total OHM velocity span can exceed 1000~km~s$^{-1}$ but typical Zeeman splitting detections have been associated with lines that have velocity widths $<$20~km~s$^{-1}$ and flux densities $>$3~mJy. The Arecibo detections required an rms flux density of 3~mJy in 0.5~km~s$^{-1}$ channels.  SKA-Mid AA4 will be able to survey all southern galaxies for OHM Zeeman splitting; a 1$''$~beam at 1.6~GHz will reach 1~mJy sensitivity with 0.5~km~s$^{-1}$ channels in just over 1~h using AA4 (3.6~h with AA$^{*}$), sufficient to detect analogues of the sources in the Arecibo sky.  If all SKA-Mid-visible Arecibo targets were searched with this threefold improvement in sensitivity, we could expect to double the number of galaxies with Zeeman detections in 2~days of observing with AA4.  The southern sky has not been thoroughly searched for OHMs; blind 21~cm surveys with SKA-Mid are bound to catalog the southern OHM population.  Deep follow-up observations of the southern OHM galaxies will yield at least another doubling of the number of Zeeman-detected galaxies; sampling the 100 brightest OHMs would require 120~h of AA4 time.  The most distant Zeeman detections occurred in OHMs at $z \sim 0.2$.  The sensitivity increase afforded by the full SKA would allow Zeeman splitting to be probed in OHMs out past $z = 1$, making these beacons invaluable tools for studying galactic magnetic fields through cosmic time.

Studies of OHM variability have relied on archival VLBI images drawn from heterogeneous observations by different teams at widely separated epochs, rather than on dedicated multiepoch monitoring campaigns; because the intervals between such observations are comparable to the variability timescales of compact masing clouds, this approach adds uncertainty when comparing results directly and testing for variability in a controlled way. The full SKA would reach this same 1~mJy rms threshold per 50~mas beam in 0.5~km~s$^{-1}$ channels in about 1~min, simultaneously delivering high spectral and spatial resolution in a single observation rather than across heterogeneous archival data. The full SKA would also extend its reach to more highly redshifted OH masers and possibly to 22~GHz water masers, which trace material orbiting black holes and jet-ISM interactions on sub-parsec scales. However, the splitting coefficient for H$_2$O is 1000 times weaker than that for OH; consequently, Zeeman splitting in water megamasers has to date eluded detection \citep{modjaz+05}.


\subsection{\HI Absorption in Damped \texorpdfstring{Ly$\bm{\alpha}$}{Ly-alpha} Absorbers}

Damped Ly$\alpha$ (DLA) systems are a class of quasar absorbers in which hydrogen remains mostly neutral. The neutral gas content of the Universe is dominated up to redshift 5 by DLAs, and the \HI layers producing the absorption are considered to be the progenitors of modern galaxies.  DLAs are perhaps the only sample of an interstellar medium in the high-redshift Universe \citep{wolfe+05}.  Accordingly, the possibility of measuring Zeeman splitting in the 21~cm line absorption in these systems would allow us to test the role of magnetic fields in galaxy formation and evolution, and constrain dynamo models for the generation and amplification of magnetic fields in the early Universe.  \citet{wolfe+08,wolfe+11} describe GBT observations of a DLA at $z=0.692$ toward 3C~286.  No Zeeman signature was detected in the Stokes~$V$ spectrum at 839.40~MHz down to a field limit of 17~$\mu$G.  Using the SKA sensitivity calculator at 839~MHz with a spectral resolution of 1.2~km~s$^{-1}$ and $R=-1$, we estimate that AA4 would require just 2~h to probe down to 5~$\mu$G field strengths in this DLA (6~h for AA$^{*}$).  The ability to probe such systems with greater sensitivity, and thereby place tighter limits on the magnetic field strength, is tantalizing.  Because of the large redshifts of these systems, all observations would be carried out in Bands~1 and~2.  These DLAs are viable targets for all phases of SKA-Mid deployment.


\section{Conclusions}

Measurement of the Zeeman effect is sensitivity-limited.  Our ability to use this method to probe magnetic fields in the Milky Way and beyond is entirely dependent on the development of next-generation telescopes that will surpass the sensitivity of any currently available observatory.  SKA-Mid will substantially improve our ability to measure Zeeman splitting in spectral lines, both in emission and absorption, providing direct measurements of the magnetic field strength and direction in atomic and molecular clouds in our own Milky Way and in external galaxies, and upper limits on the field strength along sightlines where the field remains undetected.

Within the solar system, SKA-Mid will improve measurements of cometary magnetic fields via Zeeman splitting of OH maser lines, probing the solar and interplanetary field at heliocentric distances of 1 to a few~au.  In the Milky Way, SKA-Mid Zeeman measurements will probe the magnetic field \textit{in situ} in the warm and cold neutral components of the interstellar medium, complementing the extensive SKA Faraday rotation programs aimed at probing the field in the ionized components; RRLs extend this reach to the ionized medium directly.  The first census of CNM magnetic fields throughout the previously unexplored southern sky will be readily achievable with AA4, building on a recent demonstration that Zeeman measurements and \HI filament morphology may be combined to constrain the total field strength.  Zeeman observations of HISA will map the field in cold filamentary structures, with a complete map of features such as the R-C cloud awaiting a full SKA buildout.  Diffuse 21~cm emission---seen in every direction in the sky---offers an enticing target; the SKA-Mid dishes have been designed to satisfy the Mizuguchi--Dragone condition, which should minimize the beam squint that has thus far prevented interferometric Zeeman studies of diffuse emission.  Zeeman observations of \HI in HVCs will reveal the large-scale magnetic field threading the disk--halo interface and open a new window on the role of magnetic fields in regulating gas accretion onto the Galactic disk; Zeeman measurements at the boundaries of the Fermi Bubbles will directly test whether the field has been amplified or reoriented by the expanding outflow and constrain models of bubble formation.  RRLs, stackable across hundreds of transitions in Bands~1 and~2, open the possibility of Zeeman studies in \HII regions and PDRs.  An SKA-MAGMO blind survey for OH masers in the Galactic plane will substantially expand the sample of Zeeman detections in HMSF regions, directly tracing the large-scale Galactic magnetic field in the spiral arms.

In external galaxies, the full SKA would survey the Local Group for OH maser emission and Zeeman splitting, complementing existing Faraday rotation maps of the LMC and SMC.  The improved sensitivity of SKA-Mid AA4 will enable Zeeman studies of OH kilomasers in nearby starburst galaxies such as NGC~253, NGC~4945, and NGC~5253---systems currently inaccessible to single-dish programs.  For OH megamasers, AA4 will survey the entire southern sky with a sensitivity three times better than Arecibo, doubling the number of galaxies with Zeeman detections within days of observing; the full SKA would extend this reach to OHMs at $z>1$, making these beacons invaluable probes of magnetic fields through cosmic time.  Finally, DLA absorbing systems represent perhaps the only available sample of an interstellar medium at cosmological redshifts; AA4 can probe these systems to field limits of 5~$\mu$G in just 2~h, well below the 17~$\mu$G limit of existing GBT observations, opening a direct window on the role of magnetic fields in galaxy formation and the amplification of cosmic magnetic fields in the early Universe.













\bibliographystyle{abbrvnat-maxbibnames4}
\newcommand{\actaa}{Acta Astron.} 
\newcommand{\araa}{ARA\&A} 
\newcommand{\aar}{A\&ARv} 
\newcommand{\aapr}{A\&ARv} 
\newcommand{\ab}{Astrobiol.} 
\newcommand{\aj}{AJ} 
\newcommand{\apj}{ApJ} 
\newcommand{\apjl}{ApJL} 
\newcommand{\apjs}{ApJSS} 
\newcommand{\ao}{Appl. Opt.} 
\newcommand{\apss}{Astro. \& Space Sci.} 
\newcommand{\aap}{A\&A} 
\newcommand{\aaps}{A\&AS.} 
\newcommand{\baas}{Bull. Am. Astron. Soc.} 
\newcommand{\caa}{Chinese A\&A} 
\newcommand{\cjaa}{Chinese J. A\&A} 
\newcommand{\cqg}{Class. Quantum Gravity} 
\newcommand{\gal}{Galaxies} 
\newcommand{\gca}{Geo. Cosmo. Acta} 
\newcommand{\icarus}{Icarus} 
\newcommand{\jcap}{JCAP} 
\newcommand{\jgr}{J. Geophys. Res.} 
\newcommand{\jgrp}{J. Geophys. Res. Planets} 
\newcommand{\jqsrt}{J. Quant. Spectrosc. Radiat. Transf.} 
\newcommand{\memsai}{Mem. SAIt} 
\newcommand{\mnras}{MNRAS} 
\newcommand{\nat}{Nature} 
\newcommand{\nastro}{Nat. Astron.} 
\newcommand{\ncomms}{Nat. Commun.} 
\newcommand{\nphys}{Nat. Phys.} 
\newcommand{\na}{New Astron.} 
\newcommand{\nar}{New Astron. Rev.} 
\newcommand{\physrep}{Phys. Rep.} 
\newcommand{\pra}{Phys. Rev. A} 
\newcommand{\prb}{Phys. Rev. B} 
\newcommand{\prc}{Phys. Rev. C} 
\newcommand{\prd}{Phys. Rev. D} 
\newcommand{\pre}{Phys. Rev. E} 
\newcommand{\prx}{Phys. Rev. X} 
\newcommand{\prl}{Phys. Rev. Let.} 
\newcommand{\psj}{Planet. Sci. J.} 
\newcommand{\planss}{Planet. Space Sci.} 
\newcommand{\pnas}{Proc. Natl Acad. Sci. USA} 
\newcommand{\procspie}{Proc. SPIE} 
\newcommand{\pasa}{PASA} 
\newcommand{\pasj}{PASJ} 
\newcommand{\pasp}{PASP} 
\newcommand{\rmxaa}{RMXAA} 
\newcommand{\sci}{Science} 
\newcommand{\sciadv}{Sci. Adv.} 
\newcommand{\solphys}{Sol. Phys.} 
\newcommand{\sovast}{Soviet Ast.} 
\newcommand{\ssr}{Space Sci. Rev.} 
\newcommand{\uni}{Universe} 

\bibliography{Robishaw_AASKAII_Diffuse_Zeeman} 

\end{document}